\documentclass{article}
\usepackage{cite}
\usepackage{amsmath,amssymb,amsfonts}
\usepackage{algorithmic}
\usepackage{graphicx}
\usepackage{subcaption}
\usepackage{hyperref}
\usepackage{cleveref}
\usepackage{tikz}
\usepackage{siunitx}

\def\BibTeX{{\rm B\kern-.05em{\sc i\kern-.025em b}\kern-.08em
    T\kern-.1667em\lower.7ex\hbox{E}\kern-.125emX}}

\newcommand{\Sp}[2]{\ensuremath{S_{#1#2}}}

\begin{document}

\title{A Comparative Study of Antennas for Microwave Intracranial Haemorrhage Detection Based on Power Flow and Sensitivity Metrics}
\author{August Ekman, Mikael Persson, Andreas Fhager
\thanks{\textit{Corresponding author: August Ekman.}}
\thanks{The authors are with the Department of Electrical Engineering, Chalmers University of Technology,  SE-412 96 Gothenburg, Sweden. Mikael Persson is also with the Department of Clinical Neuroscience, the Institute of Neuroscience and Physiology, University of Gothenburg, SE-413 45 Gothenburg, Sweden (email: august.ekman@chalmers.se; mikael.persson@chalmers.se; andreas.fhager@chalmers.se). }
}

\maketitle

\begin{abstract}
Microwave diagnostics has shown promise for applications such as intracranial haemorrhage detection, but system performance is often limited by multipath interference caused by electromagnetic energy propagating around, rather than through, the target tissue.
Existing antenna evaluation methods based primarily on S-parameters provide limited information about this phenomenon. This work introduces the Power Flow Ratio (PFR), a simulation-based metric that quantifies the proportion of transmitted power passing through the target volume relative to the total transmitted power.
Four antennas intended for microwave stroke diagnostics were investigated: a monopole antenna with lossy gel backing, an open ridged-waveguide antenna, a dielectric rod antenna, and a dielectric-loaded Vivaldi antenna. The antennas were assessed using simulations and measurements of reflection and transmission coefficients, transmitted power, PFR, and sensitivity to haemorrhage-induced perturbations.
The results demonstrate that PFR provides information complementary to conventional antenna metrics and offers an indirect measure of the potential susceptibility to external propagation and multipath interference. Among the evaluated designs, the open ridged-waveguide antenna exhibited the most favourable overall performance for intracranial haemorrhage detection.
\end{abstract}


\section{Introduction}
Microwave diagnostics is a rapidly growing field, with an expanding range of medical applications under investigation, including the diagnosis of chest trauma \cite{Seidel2025}, breast cancer \cite{Wang2023} and stroke \cite{Fhager2018, Guo2023, Rodriguez-Duarte2021, Liu2023, Xu2020}. A microwave diagnostic system typically consists of an array of antennas, measurement hardware, and signal-processing or image reconstruction algorithms used to analyze the acquired data and support clinical decision-making. Therefore, antenna design is a critical aspect of system development that directly influences measurement quality and diagnostic performance. A wide range of antennas have been developed and evaluated, with several review articles summarizing the different antenna types and their characteristics \cite{Shao2025, Meaney2012,Rafique2022,Misilmani2020}.

The development of antennas and antenna arrays for microwave diagnostic systems involves several considerations whose relative importance depends on the intended application. Key considerations include the operating frequency range, received signal strength, size, weight, and other practical design constraints. In addition to these factors, multipath interference can significantly affect system performance  \cite{Meaney2012}. Multipath interference arises from signal components that propagate outside of the body before reaching the receiving antenna, thereby obscuring diagnostically relevant information carried by signals that have interacted with the tissue under investigation. Therefore, antenna designs should maximize the energy coupling into the body, while minimizing external propagation paths.

Background subtraction is one of the most widely used techniques for mitigating unwanted signal components. Measurements acquired in the presence and absence of the object under test are subtracted, and the resulting differential signal is used as input to the subsequent analysis algorithm \cite{Klemm2009}. However, multipath components are often only partially removed because the object under investigation alters the propagation environment itself. Consequently, reflections, surface waves, and antenna coupling mechanisms differ between the calibration and measurement states, leaving residual multipath contributions after subtraction. These residual signals can obscure weak target responses and degrade imaging performance \cite{Meaney2012}.

Other strategies have instead been proposed to mitigate multipath interference in microwave imaging systems, including directional and shielded antennas \cite{Mobashsher2016} and time-gating techniques \cite{Fear2003}. Meaney and co-workers have emphasized the importance of minimizing multipath propagation at the hardware level through the use of a lossy matching liquid, immersing the antennas and the object under test \cite{Meaney2012,Meaney2018,Grzegorczyk2012}. In our previous work, we demonstrated improved imaging consistency by applying a lossy gel to the back of the antennas while maintaining a dry antenna-to-object interface \cite{laura-antenna}. The gel attenuates surface waves and unwanted propagation paths, reducing, although not completely eliminating, the influence of multipath signals on the reconstructed images.

In summary, the multipath interference can have a significant impact on the system performance and is a consequence of multiple design choices relating to antennas, antenna arrays, measurement hardware and coupling media. Despite extensive research into microwave diagnostic systems in several application areas, \cite{2020-cancer-review,2021-torso-review,Guo2023,Fhager2018,DeJesusArago2024} the effect of multipath interference remains difficult to quantify. Objective comparisons of different system designs are difficult, and the selection of optimal design parameters remains largely experience-based. This highlights the need for quantitative metrics that enable systematic evaluation and comparison of multipath interference across different microwave imaging systems. 

Near-field radiation patterns and the front-to-back ratio (FBR) have been used to characterize antenna directionality in microwave diagnostic systems. Directional radiation has been associated with improved image focusing and reduced artifacts \cite{Mobashsher2016}, while high FBR has been used to promote radiation towards the imaging region and reduce multipath reflections \cite{chen2023}.
Another metric proposed in \cite{Fiser2022} is antenna efficiency, defined as the ratio of the power absorbed by the phantom to the total input power.
These measures characterize antenna directionality rather than the relative distribution of power propagating through and around the object under investigation.

This study proposes the \textit{power flow ratio} (PFR), a simulation-derived metric that quantifies the fraction of electromagnetic power propagating through the target volume rather than through the surrounding region. The metric is intended to complement conventional antenna system parameters by providing an indirect indication of the potential for unwanted external propagation and multipath interference. Four representative antenna designs are compared for microwave-based intracranial haemorrhage detection using reflection and transmission coefficients, PFR, total transmitted power, and sensitivity to simulated haemorrhages.

The evaluated designs comprised a monopole antenna with lossy gel backing \cite{laura-antenna}, an open-ridged waveguide (ORWG) antenna \cite{morteza-antenna}, and a self-grounded bow-tie antenna with a dielectric waveguide \cite{moein-antenna}, all previously developed and used by our research group. A Vivaldi antenna, widely used in microwave diagnostic systems, was included as a reference case \cite{vivaldi-antenna-zhang}.

The rest of the paper is organized as follows: 
In \cref{sec:method} the measurements, simulations, and proposed metrics are described together with the antennas. The results are presented in \cref{sec:results}. \Cref{sec:discussion} discusses and compares the results in relation to the proposed metric and intracranial haemorrhage detection. \Cref{sec:conclusions} presents the conclusions.

\section{Method}
\label{sec:method}
This section presents the methods used to evaluate antenna performance with a particular focus on multipath interference and the PFR-metric. 

The objective was to compare four antenna designs for microwave-based detection of intracranial haemorrhage. The evaluation combined experimental measurements and full-wave electromagnetic simulations. Three antennas were available as physical prototypes, developed and used in our own research, and were therefore investigated using both measurements and simulations, whereas the Vivaldi antenna was evaluated solely through simulation as a reference case. 

The antennas were compared using four complementary performance metrics. First, the reflection coefficient ($S_{11}$) to assess operating bandwidth and agreement between measurements and simulations. Second, transmission measurements using receiving monopoles positioned both inside and outside a head phantom were used to investigate the radiated field distribution. Third, simulations were used to quantify how much electromagnetic energy propagated through the phantom relative to outside the phantom and how that could be quantified with the PFR-metric. Finally, simulations using a homogeneous head model containing intracranial haemorrhages were performed to assess sensitivity to internal dielectric perturbations.

\subsection{Phantom Design and Measurements}
\label{sec:phantom-n-meas}

\begin{figure}
  \centering
  \includegraphics{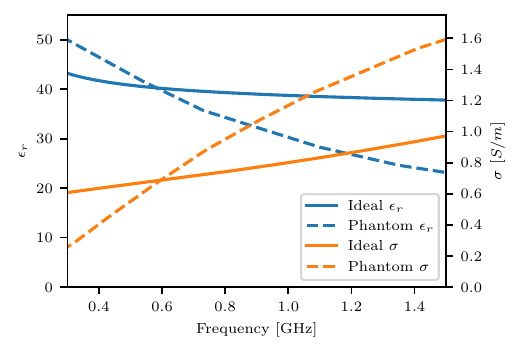}
  \caption{Permittivity and conductivity for ideal calculated average of head and for a 80\% glycerin-water mixture.
  The properties for the glycerol-water mixture were obtained from \cite{glycerin-permittivity}.}
  \label{fig:phantom-permittivity}
\end{figure}

For the measurements, a head phantom was fabricated as a homogeneous gel block measuring $19 \times 15 \times \qty{14}{cm^3}$, corresponding approximately to the dimensions of an average adult head, \cite{measure-of-man-and-woman}. The phantom consisted of a water–glycerol mixture containing mass fractions of 80 \% glycerol and 20 \% water. The composition was selected to reproduce the average dielectric properties of the human head, which was derived by taking the volume-weighted average of the properties of the heads of the Duke and Ella voxel models reported in \cite{itis-voxel-models}.
The properties of individual tissues were obtained from \cite{itis-materials}.
Figure \ref{fig:phantom-permittivity} compares the resulting water-glycerol phantom properties (dashed lines) with the ideal calculated target average head properties (solid lines).
The phantom properties does have differing frequency dependence compared to the target, however the properties are sufficiently close to enable a general comparison of the antenna performances.

Reflection and transmission measurements were performed using a Rohde \& Schwarz ZNBT8 24-port vector network analyzer (VNA) over the frequency range 0.3 - 1.5 GHz.

For transmission measurements, plain monopole antennas were used as receivers. Three receiver positions were considered: at the center of the phantom, on the side of the phantom, and on the opposite side relative to the transmitting antenna, as illustrated in \cref{fig:phantom-setup}. The monopoles outside (1 and 2) and inside (3) had physical lengths of \qty{52}{mm} and \qty{21}{mm} respectively. The internal monopole was coated with a dielectric layer to permit insertion into the phantom. The dimensions of the receiving monopoles were selected to provide resonance frequencies covering the measurement bandwidth, and their measured reflection coefficients are shown in \cref{fig:rec-monopole-S11}.

\subsection{Numerical Simulations}
\label{sec:simul}
All simulations were performed using CST Microwave Studio, Release Version 2024.05, with the finite element method (FEM) frequency domain solver.

The phantom dimensions, antenna geometries and measurement setups were replicated as closely as possible in the simulation models. The dielectric properties assigned to the phantom corresponded to the same glycerol-water mixture used for the physical phantom shown in \cref{fig:phantom-permittivity} (dashed lines), with values taken from \cite{glycerin-permittivity}. The same dielectric properties were used throughout the simulation study unless otherwise stated.

\begin{figure}
  \centering
  \includegraphics{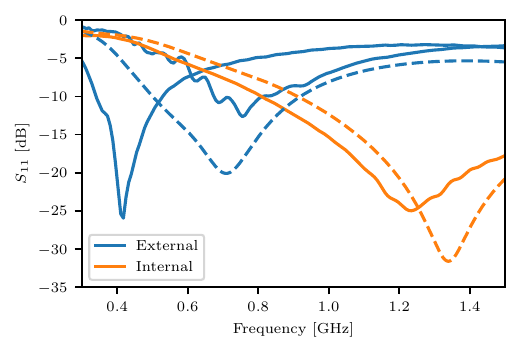}
  \caption{Measured and simulated \Sp11-parameters for the receiving monopole antennas. The solid and dashed lines indicate the measured and simulated results respectively.
  The external antenna demonstrated a very inconsistent behavior so two different measurement are shown, representing the highest and lowest observed resonance frequency.
  }
  \label{fig:rec-monopole-S11}
\end{figure}

To improve agreement between measurements and  simulations, adjustments of the lengths of the simulated monopole models were made to account for practical uncertainties in the dielectric loading.
The external monopole retained a length of \qty{52}{mm} and was positioned \qty{0.6}{mm} from the phantom surface. The internal monopole length was increased from \qty{21}{mm} to \qty{25.5}{mm}.
The resulting reflection coefficients are shown in \cref{fig:rec-monopole-S11}.
For the external monopoles, the resonance frequency was very inconsistent between different measurements due to the high sensitivity to the exact proximity to the phantom.
Therefore the antenna positioning in the simulations was chosen for a resonance in between the most extreme observed cases.
It was however found that this variation did not significantly affect the simulated transmission coefficients.

\subsection{Antenna Evaluation Metrics}

\subsubsection{Reflection Coefficient}
The first comparison metric was the reflection \Sp11, measured with the antennas adjacent to the phantom. The reflection coefficient provides information regarding impedance matching and operating bandwidth.
\Sp11 can readily be obtained in both measurements and simulations and therefore serves as an important validation metric for verifying the accuracy of the simulation models with respect to the measurements.

\subsubsection{Transmission Measurements}
\begin{figure}
  \centering
  \begin{subfigure}{0.3\textwidth}
    \centering
    \includegraphics[width=.7\textwidth]{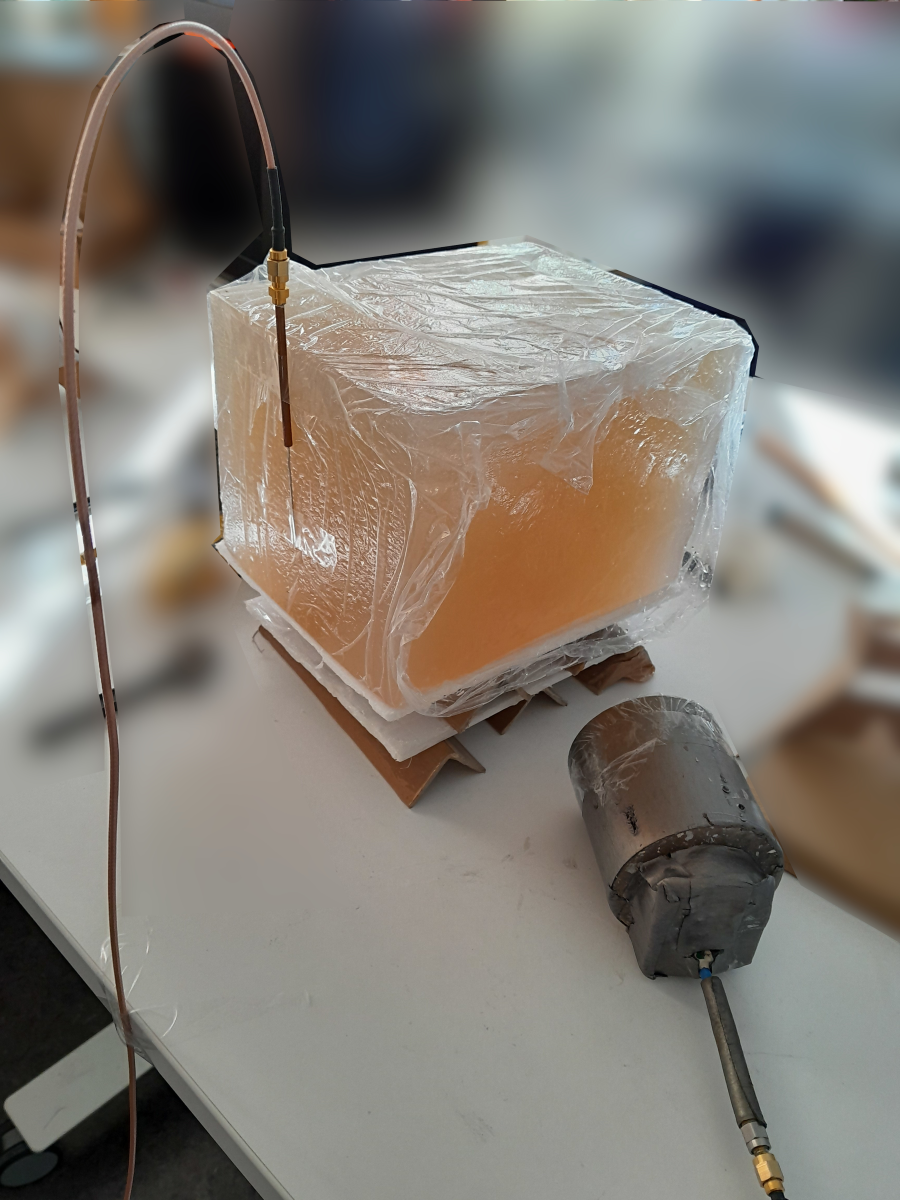}
  \end{subfigure}
  \hfill
  \begin{subfigure}{0.3\textwidth}
    \centering
    \begin{tikzpicture}
      \node[anchor=south west, inner sep=0] (image) at (0,0){%
        \includegraphics[width=\textwidth]{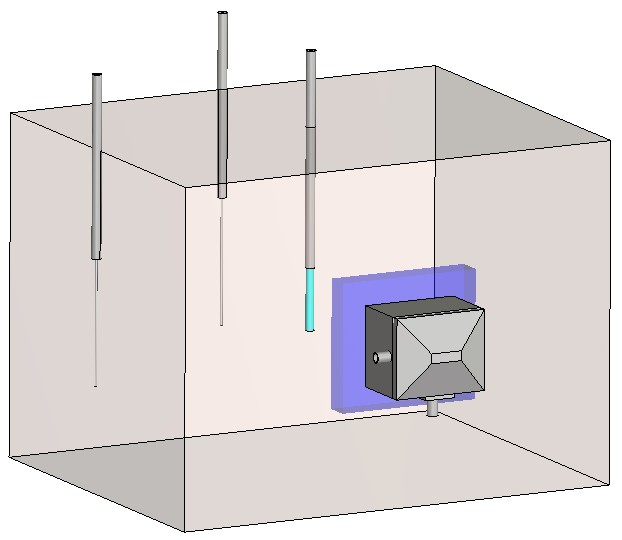}
      };
      \begin{scope}
        \path (.35\textwidth,.85\textwidth) node[above]{1};
        \path (.15\textwidth,.75\textwidth) node[above]{2};
        \path (.5\textwidth,.78\textwidth) node[above]{3};
      \end{scope}
    \end{tikzpicture}
  \end{subfigure}
  \hfill
  \begin{subfigure}{0.3\textwidth}
    \centering
    \begin{tikzpicture}
      \node[anchor=south west, inner sep=0] (image) at (0,0){%
        \includegraphics[width=\textwidth]{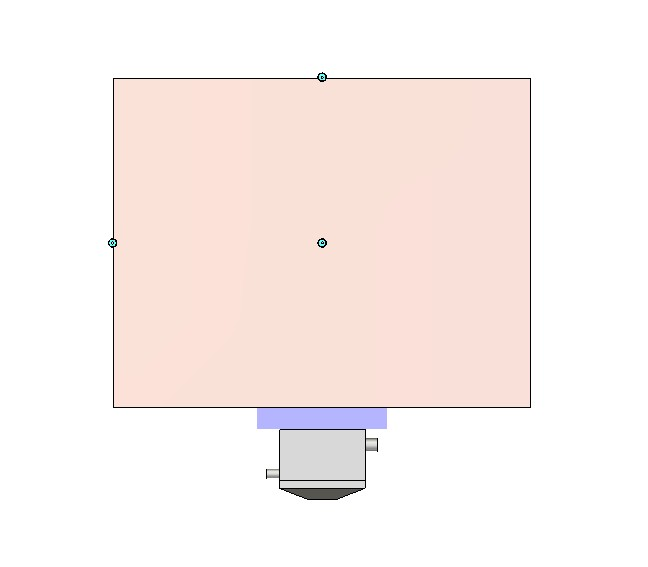}
      };
      \begin{scope}
        \path (.49\textwidth,.75\textwidth) node[above]{1};
        \path (.18\textwidth,.51\textwidth) node[above right]{2};
        \path (.49\textwidth,.51\textwidth) node[above right]{3};
        \draw[red] (0\textwidth,.5\textwidth) -- (1\textwidth,.5\textwidth);
      \end{scope}
    \end{tikzpicture}
  \end{subfigure}
  \caption{Measurement setup and corresponding model for simulation. Positions 1, 2, and 3 correspond to the opposite, side, and center receivers, respectively. The red line marks the location of the plane over which the power flow will be analyzed.
  }
  \label{fig:phantom-setup}
\end{figure}

To investigate the distribution of radiated energy, transmission measurements were performed between the transmitting antenna and the receiving monopoles shown in \cref{fig:phantom-setup}. Measurements were obtained for receivers positioned at the center, side, and opposite locations. Corresponding simulations were also performed using the same geometry and antenna placement.

Unlike the reflection coefficient, which only quantifies the power accepted by the antenna, transmission coefficients provide information about where the radiated energy has propagated. By using the same receiving antennas for all measurements, differences in received signal strength can be attributed primarily to the transmitting antennas. Furthermore, the omnidirectional receiving patterns of the monopoles enable direct comparison between antennas with different radiation characteristics and patterns.


\subsubsection{Power Flow Ratio}

The proposed power flow ratio (PFR) was defined as the ratio of the electromagnetic power flowing inside the phantom to the total power flowing both inside and outside the phantom. The power was evaluated across a plane intersecting the center of the phantom. Although the position of this plane is arbitrary to some extent, the center plane provides a simple and reproducible reference for comparing different antenna configurations. By applying the same integration plane and phantom geometry to all antennas, the PFR enables a relative assessment of how effectively the electromagnetic power is directed through, rather than around, the phantom. 

As illustrated in \cref{fig:phantom-setup}, the plane was positioned halfway between the transmitter and opposite receiver, and oriented perpendicular to the line connecting transmitting and the opposite receiving antenna. The time-averaged power crossing this plane was integrated separately over the regions inside and outside the phantom.
The power flow ratio was defined as
\begin{equation}
  \text{PFR}(f) = \frac{P_\text{inside}(f)}{P_\text{inside}(f) + P_\text{outside}(f)}
  = \frac{P_\text{inside}(f)}{P_\text{total}(f)}
\end{equation}
where ${P_{inside}}$ and ${P_{outside}}$ denote the time-averaged power crossing the regions of the evaluation plane inside and outside the phantom, respectively. The power crossing the region inside the phantom was calculated from the normal component of the time-averaged Poynting vector as
\begin{equation}
P_\mathrm{inside}(f)=\int_{A_\mathrm{inside}} \langle\mathbf{S}(f)\rangle \cdot \mathbf{\hat{n}}\; dA
\end{equation}
where $\langle\mathbf{S}(f)\rangle$ is the time-averaged Poynting vector, $\mathbf{\hat{n}}$ is the unit normal to the evaluation plane, and $A_{\mathrm{inside}}$ denotes the portion of the plane intersecting the phantom. The outside power, $P_{\mathrm{outside}}$, was calculated analogously by integrating over the region of the plane outside the phantom.

The PFR therefore quantifies the fraction of power crossing the evaluation plane that propagates through the phantom rather than through the surrounding region. External propagation may contribute to surface-wave and multipath components, and the PFR thus provides an indirect measure of the potential susceptibility of an antenna configuration to such propagation.

\subsubsection{Homogeneous Head Model Simulations}

To evaluate performance under more realistic imaging conditions,  simulations were conducted using a head model from the CST component library \cite{cst}. The model is homogeneous and in a non-voxel format, suitable for FEM simulation.

\begin{figure}
  \centering
  \begin{subfigure}[b]{0.25\textwidth}
    \centering
    \includegraphics[width=\textwidth]{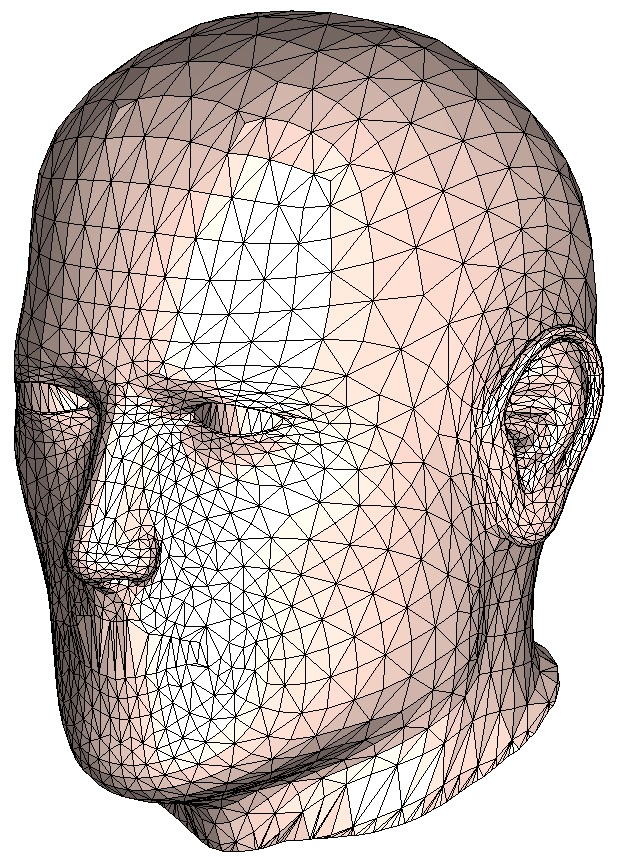}
  \end{subfigure}
  \begin{subfigure}[b]{0.25\textwidth}
    \centering
    \begin{tikzpicture}
      \node[anchor=south west, inner sep=0] (image) at (0,0){%
        \includegraphics[width=\textwidth]{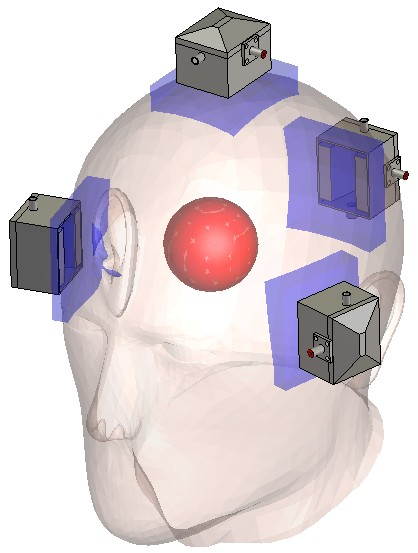}
      };
      \begin{scope}[xscale=\textwidth/1cm, yscale=\textwidth/1cm]
        \path (.8,.35) node{1};
        \path (.1,.55) node{2};
        \path (.87,1.1) node{3};
        \path (.37,1.2) node{4};
      \end{scope}
    \end{tikzpicture}
  \end{subfigure}
  \begin{subfigure}[b]{0.25\textwidth}
    \centering
    \begin{tikzpicture}
      \node[anchor=south west, inner sep=0] (image) at (0,0){%
        \includegraphics[width=\textwidth]{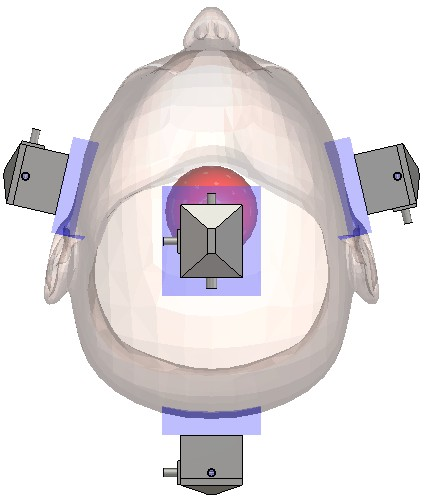}
      };
      \begin{scope}
        \path (.12\textwidth,.9\textwidth) node{1};
        \path (.9\textwidth,.9\textwidth) node{2};
        \path (.35\textwidth,.1\textwidth) node{3};
        \path (.35\textwidth,.6\textwidth) node{4};
      \end{scope}
      \draw[<->] (.165\textwidth,0) -- (.835\textwidth,0) node[midway, below] {\qty{15}{cm}};
      \draw[<->] (\textwidth,.19\textwidth) -- (\textwidth,1.06\textwidth) node[midway, right] {\qty{20}{cm}};
    \end{tikzpicture}
  \end{subfigure}
  \caption{The homogeneous head model illustrated with one of the antenna types in position. The same antenna positions were used for all antenna types. A haemorrhage is shown as a red sphere in the center of the head model.}
  \label{fig:sam-setup}
\end{figure}

Four antennas of the same type were positioned around the head model as illustrated in \cref{fig:sam-setup}.
The head volume was assigned the dielectric properties corresponding to the average head properties that were calculated when deciding on the water-glycerol ratio used for the phantom.
The properties that we use for the homogeneous head simulations are shown in \cref{fig:phantom-permittivity}, there marked as realistic.

\begin{figure}
  \centering
  \includegraphics{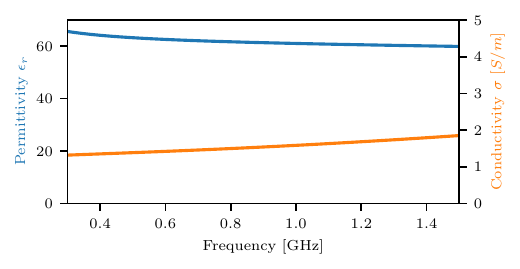}
  \caption{Permittivity and conductivity of blood \cite{itis-materials}.}
  \label{fig:blood-permittivity}
\end{figure}
\begin{figure}
  \centering
  \begin{subfigure}[b]{0.3\columnwidth}
    \centering
    \includegraphics[width=\textwidth]{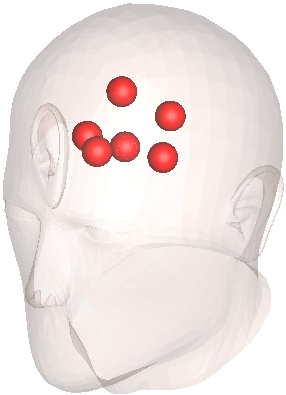}
  \end{subfigure}
  \hfill
  \begin{subfigure}[b]{0.3\columnwidth}
    \centering
    \includegraphics[width=\textwidth]{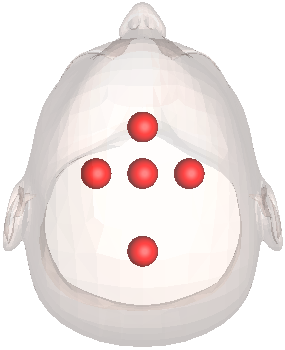}
  \end{subfigure}
  \hfill
  \begin{subfigure}[b]{0.3\columnwidth}
    \centering
    \includegraphics[width=\textwidth]{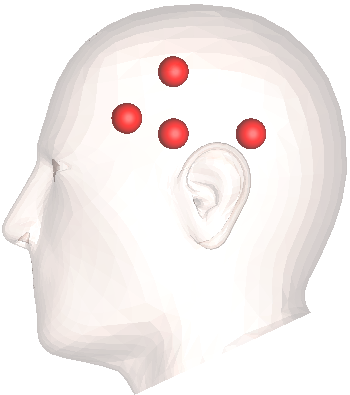}
  \end{subfigure}
  \caption{Location of the 6 simulated haemorrhages in homogeneous head model.
  Here the haemorrhage size is reduced to better highlight the locations.
  In the top-facing and side-facing images some locations obstruct each other so that not all of them are visible.
  }
  \label{fig:bleeding-locations}
\end{figure}
Intracranial haemorrhages were modeled as spherical inclusions with a diameter \qty{50}{mm} corresponding to a volume of approximately \qty{65}{ml}.
This size was chosen as a compromise between being large enough to give results easily seen in simulations for all antennas and small enough to think of as points when assessing the antenna sensitivity to various locations in the head model.
Only one size was modeled since that was deemed sufficient for comparing the antennas.
The dielectric properties of blood were obtained from \cite{itis-materials} and are shown in \cref{fig:blood-permittivity}. Six haemorrhage locations were investigated. One haemorrhage was positioned at the center of the head, whereas five additional cases were positioned closer to the skull in the forwards, backwards, left, right, and upward directions as illustrated in \cref{fig:bleeding-locations}.

For each antenna type, simulations were first performed using the homogeneous head model without a haemorrhage. Simulations were then repeated for each haemorrhage location.

Detection sensitivity was assessed by analyzing the resulting perturbations of the measured scattering parameters. Desirable antenna characteristics include strong transmission through the head combined with a large variation in signal amplitude in the presence of a haemorrhage. The comparison therefore provides an indication of the suitability of each antenna for microwave-based haemorrhage detection in the head.

\subsection{Antenna Models}
\label{sec:method-antennas}

\begin{figure}[h!]
  \centering
  \begin{subfigure}{0.2\textwidth}
    \centering
    \includegraphics[width=\textwidth]{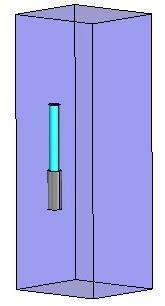}
    \caption{}
  \end{subfigure}%
  \vrule%
  \hspace{2pt}%
  \begin{subfigure}{0.2\textwidth}
    \centering
    \begin{subfigure}{\textwidth}
      \centering
      \includegraphics[width=\textwidth]{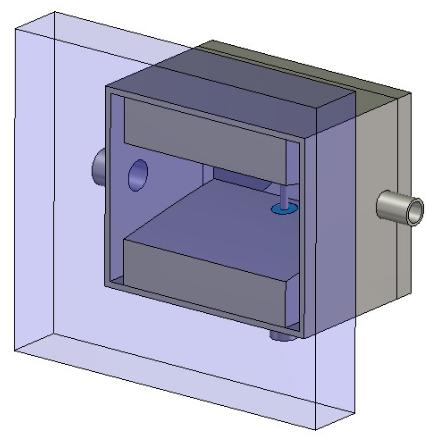}
    \end{subfigure}
    \vfill
    \begin{subfigure}{\textwidth}
      \centering
      \includegraphics[width=\textwidth]{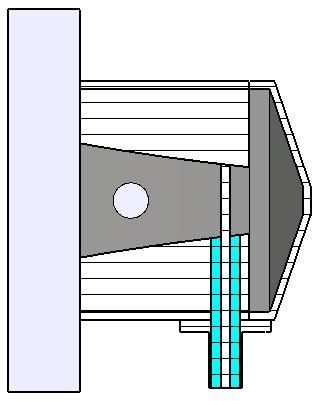}
    \end{subfigure}
    \caption{}
  \end{subfigure}%
  \hspace{2pt}%
  \vrule%
  \hspace{2pt}%
  \begin{subfigure}{0.3\textwidth}
    \centering
    \begin{subfigure}{\textwidth}
      \centering
      \includegraphics[width=\textwidth]{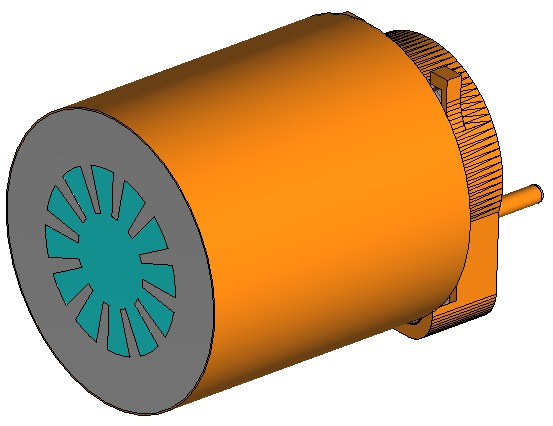}
    \end{subfigure}
    \vfill
    \begin{subfigure}{\textwidth}
      \centering
      \includegraphics[width=\textwidth]{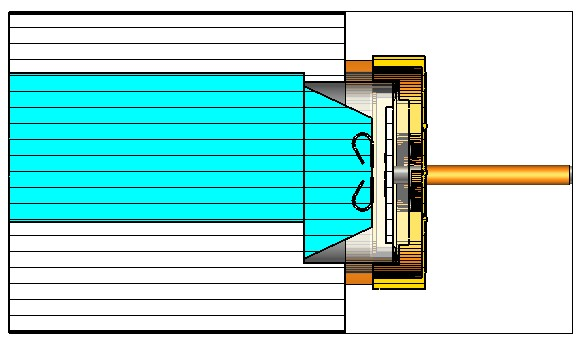}
    \end{subfigure}
    \caption{}
  \end{subfigure}%
  \hspace{2pt}%
  \vrule%
  \hspace{2pt}%
  \begin{subfigure}{0.25\textwidth}
    \centering
    \begin{subfigure}{\textwidth}
      \centering
      \includegraphics[width=\textwidth]{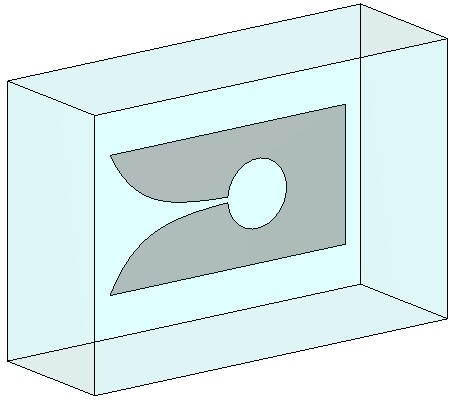}
    \end{subfigure}
    \vfill
    \begin{subfigure}{\textwidth}
      \centering
      \includegraphics[width=\textwidth]{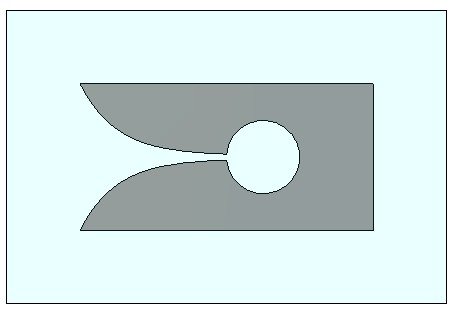}
    \end{subfigure}
    \caption{}
  \end{subfigure}
  \caption{Models of the investigated antennas.
  These are (a) the monopole antenna in a lossy gel, (b) ORWG antenna with hydrogel in blue in front, (c) DRA antenna, and (d) Vivaldi antenna.}
  \label{fig:antennas}
\end{figure}

Four antenna designs were evaluated in this study. Three antennas were developed as parts of our own research for different purposes and therefore available as physical prototypes.  They were studied using both measurements and simulations. The antennas were the lossy monopole \cite{laura-antenna}, the open ridged-waveguide (ORWG) antenna \cite{morteza-antenna}, and a dielectric rod antenna (DRA) \cite{moein-antenna}. A fourth antenna, a dielectric-loaded Vivaldi antenna, was evaluated only through simulations as a benchmark \cite{vivaldi-antenna-zhang}. \Cref{fig:antennas} shows the investigated antenna models.

\subsubsection{Lossy Monopole}
\label{sec:method-monopole-antenna}
The first antenna investigated was a monopole embedded in a lossy gel environment to suppress the fields propagating outside the body, shown in \cref{fig:antennas}. The antenna was originally developed for microwave detection of muscle ruptures and is designed to operate as part of a semi-circular antenna array surrounding the measurement object \cite{laura-antenna}.
In the current work we use a smaller version of that original antenna, with the monopole length being reduced from 35 to \qty{22}{mm} for a operating frequency range around \qty{1}{GHz}.
Prior to measurements, this antenna cavity was filled with a conductive gel consisting of saline water gelled with agar. The salt concentration was adjusted to achieve a conductivity close to $\sigma=\qty{7}{S/m}$ within the frequency range of interest. The gel was molded around the monopole element to create the final structure.

The monopole antenna was found to be highly sensitive to the dielectric properties of the surrounding gel, and to the depth of the radiating monopole inside the absorbing gel.
This indicates a strong dependence on the local antenna-body interface.
To achieve satisfactory agreement between measurements and simulations, some modifications were introduced to the simulation model. The monopole length was reduced by 23\% and the depth of the monopole element inside the gel was reduced from \qty{1.5}{mm} to \qty0{mm}. The resulting simulated reflection coefficient \Sp11 is compared with measurements in \cref{fig:S11-sim-meas}.

\subsubsection{Open Ridged-Waveguide Antenna}
The second antenna was an open ridged-waveguide (ORWG) antenna originally developed for hyperthermia treatment of cancer.
This had an operating frequency range of \qty{400}{MHz} to \qty{800}{MHz} \cite{morteza-antenna}.
The antenna structure incorporates a dielectric loading region that is normally water-cooled during hyperthermia operation. In the present work no cooling is required and the cavity was instead filled with a hydrogel made from distilled water and agar. To emulate the intended operating conditions of the ORWG antenna, the hydrogel extended \qty1{cm} in front of the antenna. This configuration provides dielectric matching to the tissue phantom and improves electromagnetic coupling. The antenna with hydrogel in blue is shown in \cref{fig:antennas}.

\subsubsection{Dielectric Rod Antenna (DRA)}
The third antenna was a self-grounded bow-tie antenna incorporating a dielectric waveguide to direct electromagnetic energy into biological tissue. In this work the antenna is referred to as the dielectric rod antenna (DRA).
The antenna was originally developed for microwave-based detection of ischemic stroke, where injected saline alters the dielectric properties of cerebral tissue. The dielectric rod is designed to improve penetration into the head while limiting undesirable surface-wave propagation.
The reported bandwidth is \qty{640}{MHz} to \qty{1530}{MHz} \cite{moein-antenna}.

Before measurements, the dielectric rod section was filled with the same hydrogel used for the ORWG antenna.

\subsubsection{Vivaldi Antenna}
The fourth antenna investigated was a dielectrically loaded Vivaldi antenna designed for operation around \qty{1}{GHz}. The antenna was enclosed within a non-conductive dielectric material having relative permittivity $\varepsilon_r = 40$ to reduce its physical dimensions and improve the coupling to tissue.
Vivaldi antennas are frequently employed in microwave imaging systems due to their wide bandwidth and directional characteristics. However, in this study it was only studied through numerical simulations as a comparative reference case.

\section{Results}
\label{sec:results}

This section presents the measured and simulated results for the antennas. The antennas were evaluated with respect to impedance matching, transmission characteristics, field distributions, and performance in a simplified head model as described in the previous section.

\subsection{Comparison of Reflection Coefficients}
\begin{figure}
  \centering
  \includegraphics{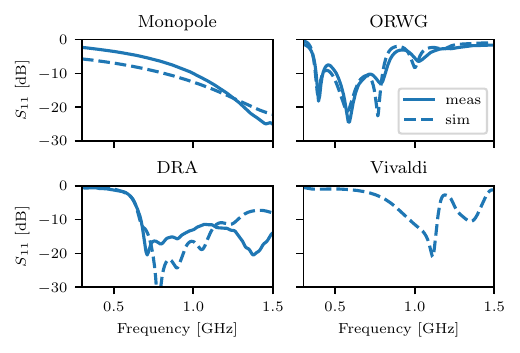}
  \caption{Measured and simulated \Sp11 for the block phantom.
    For the Vivaldi antenna, only simulation results were available.}
  \label{fig:S11-sim-meas}
\end{figure}

Measured and simulated \Sp11 responses obtained with the block shaped phantom and antenna placement as in \cref{fig:phantom-setup} are shown in \cref{fig:S11-sim-meas}. The corresponding \qty{-10}{dB} frequency ranges are summarized in \cref{tab:freq-range}.
Within these frequency ranges we have allowed for small spans where \Sp11 reaches above \qty{-10}{dB}.
This was mostly applicable for the Vivaldi antenna, and to a smaller extent the ORWG antenna.

The monopole antenna exhibited the broadest measured bandwidth. The ORWG antenna operated at the lowest frequencies and showed excellent agreement between measurements and simulations. The DRA antenna also exhibited a broad bandwidth, with good agreement near resonance but increasing discrepancies at higher frequencies. The Vivaldi antenna exhibited the narrowest bandwidth.

The measurements agree well with the simulations, and capture the trends in \Sp11 well. However, the DRA antenna shows increasing deviation between simulations and measurements at higher frequencies.
Both the simulated and measured results are consistent with the original publication \cite{moein-antenna}.

\begin{table}
  \centering
  \begin{tabular}{c | c c | c c}
    & \multicolumn{2}{c|}{Measured} & \multicolumn{2}{c}{Simulated} \\
    & $f_\text{min}$ & $f_\text{max}$ & $f_\text{min}$ & $f_\text{max}$ \\
    \hline
    Monopole & 1.00 & $> 1.50$ & 0.82 & $> 1.50$ \\
    ORWG & 0.38 & 0.81 & 0.38 & 0.80 \\
    DRA & 0.67 & $> 1.50$ & 0.66 & 1.28 \\
    Vivaldi & & & 0.96 & 1.37 \\
  \end{tabular}
  \caption{Measured and simulated antenna frequency ranges in \unit{GHz}. For some antennas, these spans can include small frequency ranges where the resonance goes above \qty{-10}{dB}.
  }
  \label{tab:freq-range}
\end{table}

\subsection{Transmission with Receiving Monopoles}

\begin{figure*}
  \centering
  \includegraphics{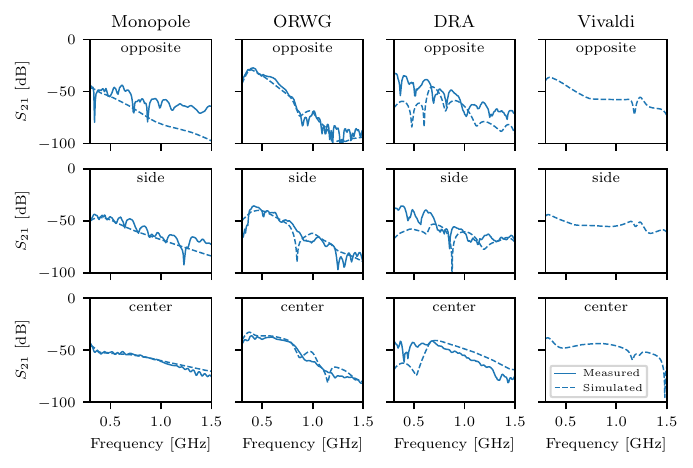}
  \caption{Measured and simulated transmission \Sp21 for block phantom with receiving monopole antennas.
    The solid and dashed lines are the measured and simulated values respectively}
  \label{fig:S21-probe-comparison}
\end{figure*}

The transmission characteristics were evaluated using measured and simulated \Sp21 parameters.
The results for all three receiver positions are shown in \cref{fig:S21-probe-comparison}.
For all antennas, transmission generally decreased with increasing frequency. The reduction was most pronounced for the monopole, ORWG, and DRA antennas. In contrast, the Vivaldi antenna exhibited a comparatively uniform transmission magnitude across its operating frequency range.
For a quantitative assessment of the discrepancy between measurements and simulations, the average difference in \Sp21 in dB was taken for each antenna and each receiving antenna position.
The average was only calculated over each antennas measured frequency range respectively as shown in \cref{tab:freq-range}, and the resulting values can be seen in \cref{tab:sim-meas-diff}.
Across all receiver positions, the ORWG antenna demonstrates the smallest average discrepancy between measured and simulated results. Measurements with both the monopole and DRA antennas capture the trends of the transmission well. Particularly the opposite receiver antenna for the monopole, and the frequencies below  \qty{0.7}{GHz} for the DRA antenna show larger discrepancies.
\begin{table}
  \centering
  \begin{tabular}{c | c c c}
    & Monopole & ORWG & DRA \\
    \hline
    Opposite & 23.8 & 2.6 & 11.0 \\
    Side & 7.5 & 2.8 & 5.1 \\
    Center & 3.0 & 1.8 & 7.8 \\
  \end{tabular}
  \caption{Average difference between measured and simulated \Sp21 in \unit{dB} for each measured antenna and receiver position. The averages were taken over each antennas respective measured bandwidth, presented in \cref{tab:freq-range}}
  \label{tab:sim-meas-diff}
\end{table}

\subsection{Power flow ratio and Field Strength}
\begin{figure}
  \centering
  \includegraphics{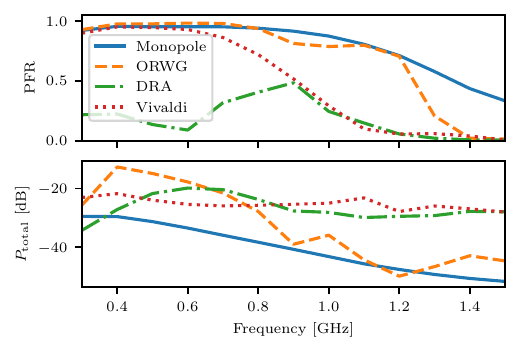}
  \caption{The upper graph shows the fraction of power inside the phantom relative to the total power.
  The lower graph shows the integrated power inside and outside the phantom on the plane in \unit{dB}.  
  }
  \label{fig:power-ratios}
\end{figure}

The simulated power flow through and around the phantom is shown in \cref{fig:power-ratios}. The figure presents both the fraction of power propagating through the phantom relative to the total transmitted power, i.e. the PFR. A PFR of zero indicates that all power crossing the evaluation plane does so outside the phantom, whereas a PFR of unity indicates that all power crossing the plane does so within the phantom. The graph also shows the total power integrated over the same plane in the center of the phantom. 

At lower frequencies, the monopole, ORWG, and Vivaldi antennas exhibited power ratios close to unity, while the ratio decreased with increasing frequency for all antennas. The monopole maintained a high power ratio but relatively low transmitted power. The ORWG antenna combined a high power ratio with high transmitted power across its operating band. The DRA antenna exhibited the lowest power ratio at lower frequencies, increasing toward the levels of the other antennas at higher frequencies. The Vivaldi antenna showed lower power ratios at higher frequencies but maintained approximately constant transmitted power across the investigated frequency range.

\subsection{Simulations of Homogeneous Head}

\begin{figure}
  \centering
  \includegraphics{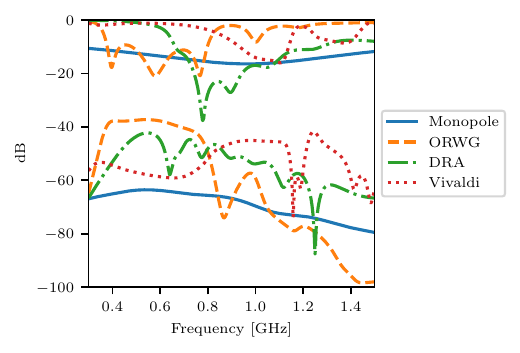}
  \caption{Simulated S11 and S21 for all antennas with left and right side of a head phantom}
  \label{fig:simulation-head-baseline}
\end{figure}

The antennas were subsequently evaluated in simulations using a homogeneous head model. Simulated \Sp11 and \Sp21 responses are shown in \cref{fig:simulation-head-baseline}, where \Sp21 corresponds to transmission between antennas located on opposite sides of the head.

The simulated \Sp11 responses generally followed the trends observed in the block phantom simulations. The largest difference was observed for the monopole antenna, whose resonance shifted toward lower frequencies in the head model.
This is consistent with the monopole antenna being the most sensitive to small variations close to the radiating element,
such as positioning relative to the phantom and the phantom dielectric properties, both of which were different between the block phantom and homogeneous head simulations.

The ORWG antenna exhibited the highest transmission levels (approximately \qty{-40}{dB}), whereas the monopole antenna exhibited the lowest (<\qty{-60}{dB}). The DRA and Vivaldi antennas produced intermediate transmission levels, generally between \qty{-60}{dB} and \qty{-50}{dB} across their operating bands.

\subsection{Sensitivity to Simulated Intracranial Haemorrhages}

\begin{figure*}
  \centering
  \includegraphics{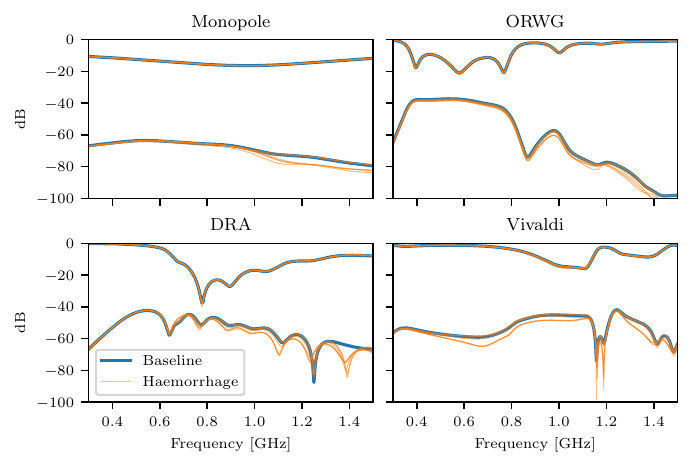}
  \caption{Comparison of \Sp11 and \Sp21 for simulated homogeneous head (baseline) and with various haemorrhages}
  \label{fig:simulation-head-compare-bleeding}
\end{figure*}

The sensitivity of the antenna configurations to intracranial haemorrhage was evaluated by introducing haemorrhagic inclusions into the homogeneous head model. The resulting responses are presented in \cref{fig:simulation-head-compare-bleeding}.
Both the reflection and transmission had a response, however in the figure the response is only visible for the \Sp21-parameters due to the higher base level for the \Sp11-parameters.

\begin{figure*}
  \centering
  \includegraphics{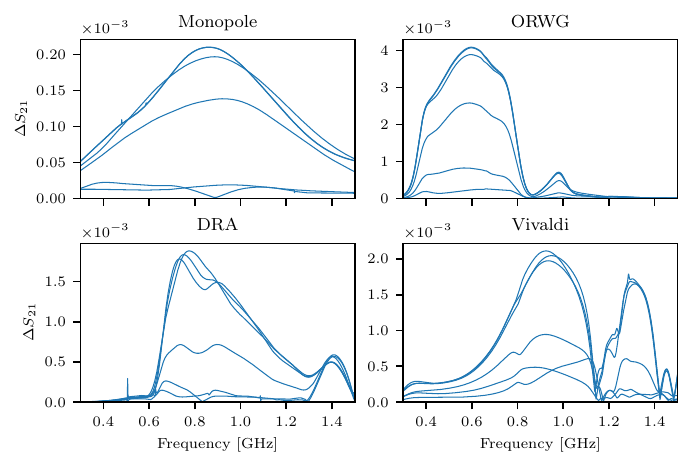}
  \caption{Absolute change in simulated \Sp21 with linear scale due to various haemorrhage scenarios for right-to-left head transmission.
  }
  \label{fig:simulation-head-compare-bleeding-linear-S21}
\end{figure*}

The largest changes in \Sp21 were observed for the DRA and Vivaldi antennas. Absolute differences in transmission are shown in \cref{fig:simulation-head-compare-bleeding-linear-S21}, where the ORWG antenna exhibited the largest absolute signal variation, followed by the DRA and Vivaldi antennas. The monopole antenna exhibited substantially smaller changes than the remaining antenna designs.

\begin{figure}
  \centering
  \includegraphics{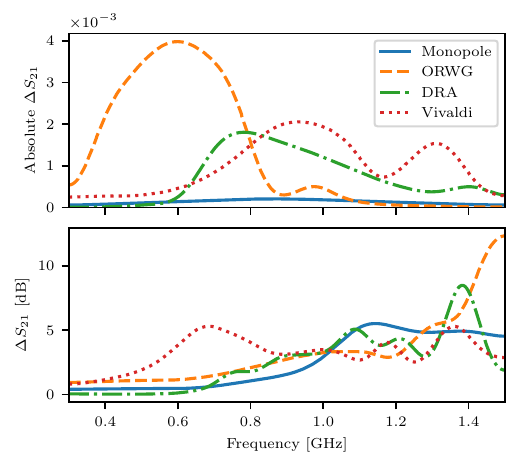}
  \caption{Maximum deviation in simulated \Sp21 from the baseline model across all haemorrhage scenarios for right-to-left transmission through the head. The upper plot shows the absolute difference, while the lower plot shows the relative difference in dB. Curves were smoothed using a Gaussian filter to highlight overall trends.
  }
  \label{fig:simulation-head-max-bleeding-variation-S21}
\end{figure}

The maximum transmission differences observed across all haemorrhage scenarios are summarized in \cref{fig:simulation-head-max-bleeding-variation-S21}. The DRA, ORWG, and Vivaldi antennas consistently exhibited larger responses to the introduced perturbations than the monopole antenna.

\begin{figure*}
  \centering
  \includegraphics{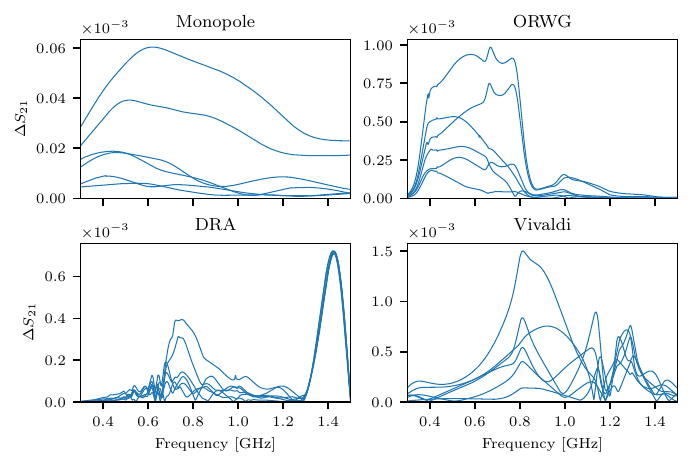}
  \caption{Absolute difference between baseline and haemorrhage-case simulated \Sp41 responses. The receiving antenna is at the top of the head.}
  \label{fig:simulation-head-compare-bleeding-linear-S41}
\end{figure*}

Additional receiver positions located at the back and top of the head exhibited similar trends. The receiver at the back of the head produced signal variations of comparable magnitude to those observed for the left-right transmission path. The receiver at the top of the head exhibited smaller transmission changes, as shown in \cref{fig:simulation-head-compare-bleeding-linear-S41}.

\section{Discussion}
\label{sec:discussion}
The investigated antennas exhibited distinct strengths and limitations with respect to impedance matching, transmission performance, sensitivity to internal perturbations, and robustness to variations in the measurement setup. In this section, these findings are discussed in the context of microwave-based diagnostic systems. Subsequently, the proposed evaluation methodology is discussed.

\subsection{Comparison of Antenna Designs}
\label{sec:discussion-results}

\subsubsection{Impedance Matching and Robustness}
The measured and simulated \Sp11 responses indicate notable differences among the investigated antennas.
Initially the monopole antenna exhibited the largest variation between phantom measurements and simulations,
as was discussed in \cref{sec:method-monopole-antenna}.
To achieve the agreement in \Sp11 shown in \cref{fig:S11-sim-meas} the simulated monopole antenna geometry first needed to be modified.
These modifications were not necessary for the ORWG and DRA antenna.
The reason is that for this monopole antenna a substantial portion of the active near field extends beyond the antenna volume and into the adjacent tissue or phantom material.
Consequently, small variations in antenna positioning, core insertion depth, contact conditions, or phantom properties can significantly affect the measured response.

In contrast, the ORWG and DRA antennas exhibited smaller variations between measurements and simulations and generally better agreement with numerical models. Their fields are concentrated further inside the antenna structures, reducing the influence of local interface variations and leading to more stable responses. The same is likely true for the Vivaldi antenna, but measurements would be needed to confirm this. 

This finding is consistent with the monopole antenna exhibiting the largest change in \Sp11 between the simulated block phantom and homogeneous head model.
These two setups have differing contacts between antenna and model since the head model has an uneven surface.
The simulated block phantom and head model also had different dielectric properties, further affecting the antenna performance.

Since variations originating from the antenna-body interface do not contain information about internal tissue properties, they may obscure diagnostically relevant signal changes. The ORWG, DRA and Vivaldi antennas therefore appear more robust than the monopole antenna for the investigated application.

\subsubsection{Transmission Performance}
Transmission strength is an important consideration since stronger received signals generally improve measurement robustness and reduce hardware requirements. In this work, the \Sp21-parameters were studied with the receiving antenna positioned on the opposite side of the phantom from the transmitting antenna. This configuration was evaluated for both the homogeneous phantom block, shown in \cref{fig:S21-probe-comparison}, and the homogeneous head phantom, shown in \cref{fig:simulation-head-baseline}.

The transmission results highlight a trade-off between operating frequency and received signal strength. The low transmission levels of the monopole may reduce sensitivity to small tissue perturbations and impose greater demands on the measurement hardware. Conversely, the high transmission obtained with the ORWG is partly associated with its lower operating frequency, where tissue attenuation is reduced.

\subsubsection{Assessment of Multipath Interference} 

One of the primary objectives of this study was to evaluate the potential influence of multipath propagation. The PFR quantifies the fraction of transmitted power propagating through the target volume rather than through the surrounding region. Because externally propagating fields may contribute to unwanted surface-wave and multipath components, the PFR provides an indirect indicator of an antenna configuration's potential susceptibility to multipath interference.

The PFR should, however, be considered together with the total transmitted power in \cref{fig:power-ratios}. A high PFR indicates that a large fraction of the transmitted energy propagates through the phantom, but does not necessarily imply a high absolute power level. Conversely, strong transmission alone does not indicate whether the energy propagates primarily through or around the target volume. The two measures therefore provide complementary information on antenna performance. This distinction is apparent in the results, where the investigated antennas exhibit different combinations of PFR and transmitted power.

The haemorrhage simulations provide a complementary assessment of whether the transmitted fields are sensitive to perturbations within the target volume. Considering PFR, transmitted power, and haemorrhage-induced signal changes together provides a more complete assessment than any of these quantities individually. The ORWG antenna shows a favorable combination of high PFR, high transmitted power, and sensitivity to the simulated haemorrhages within its operating frequency range. The DRA and Vivaldi antennas also show substantial responses to the introduced haemorrhages, whereas the monopole combines a relatively high PFR with lower transmitted power and smaller haemorrhage-induced signal changes.

These results highlight the intended role of the PFR as a complementary rather than stand-alone performance metric. Combined with transmitted power and sensitivity to internal perturbations, the PFR provides a more complete basis for comparing antenna configurations with respect to multipath susceptibility and diagnostic sensitivity.

\subsection{PFR as an Antenna Evaluation Metric}
\label{sec:discussion-comparisons}

The proposed comparisons provided a structured framework for evaluating microwave antenna systems intended for diagnostic applications. By combining impedance matching, transmission measurements, field-distribution analyses, and sensitivity studies, a more complete characterization was obtained than would be possible using any individual metric alone. In particular, the power flow ratio provided information that could not be inferred directly from S-parameters alone, namely the fraction of transmitted energy propagating through the target volume rather than around it. This offers additional insight into the potential influence of multipath propagation and the extent to which the radiated fields interact with the tissue of interest. The PFR cannot be obtained directly from the discrete transmission measurements employed in this study and was therefore evaluated numerically. The experimental S-parameter measurements were instead used to assess the ability of the numerical models to reproduce the principal antenna and propagation characteristics. Consequently, the PFR relies on the accuracy of the numerical representation of the antenna, phantom, and surrounding environment. It should also be stressed that the power flow ratio is not a pure antenna metric, rather, it depends on the antenna, object under test, coupling medium, and surrounding environment.

The methodology is not limited to head-imaging applications. For alternative diagnostic targets, such as the thorax or extremities, the phantom geometry and dielectric properties could be adapted while retaining the overall evaluation procedure. This enables systematic comparison of antenna candidates across different clinical applications.

Another important aspect is robustness to manufacturing tolerances and measurement variability. While indications of such effects were observed for the monopole antenna, a dedicated investigation would be required to quantify their impact.

Finally, practical considerations such as antenna dimensions, manufacturing complexity, mechanical robustness, and production cost may significantly influence the suitability of an antenna design for a clinical system. The relative importance of these factors will depend on the intended application and deployment scenario.

\section{Conclusions}
\label{sec:conclusions}
This paper presented a framework for evaluating antennas for microwave diagnostic applications, combining conventional antenna metrics with the proposed power flow ratio and sensitivity analyses. The methodology was applied to four antenna designs intended for intracranial haemorrhage detection, enabling comparison of impedance matching, transmission performance, robustness, and potential susceptibility to externally propagating multipath components.

The results demonstrated clear trade-offs between the investigated antenna designs. Overall, the ORWG antenna provided the most favorable combination of transmission performance, robustness, and power penetration, while the DRA and Vivaldi antennas showed competitive performance within their respective operating frequency ranges. The proposed evaluation framework provides a systematic basis for application-specific antenna selection and may facilitate more standardized comparisons between microwave diagnostic systems.

\section*{Acknowledgment}
During manuscript preparation, Microsoft Copilot was used solely to assist with language editing and improvement of readability.
All technical content, analysis, interpretations, and conclusions were developed, verified, and approved by the authors.

\bibliographystyle{plain}
\bibliography{library}

\end{document}